# The Task-oriented Queries Benchmark (ToQB)

Keun Soo YIM
*Google*
yim@google.com

**Abstract**

*Task-oriented queries (e.g., one-shot queries to play videos, order food, or call a taxi) are crucial for assessing the quality of virtual assistants, chatbots, and other large language model (LLM)-based services. However, a standard benchmark for task-oriented queries is not yet available, as existing benchmarks in the relevant NLP (Natural Language Processing) fields have primarily focused on task-oriented dialogues. Thus, we present a new methodology for efficiently generating the Task-oriented Queries Benchmark (ToQB) using existing task-oriented dialogue datasets and an LLM service. Our methodology involves formulating the underlying NLP task to summarize the original intent of a speaker in each dialogue, detailing the key steps to perform the devised NLP task using an LLM service, and outlining a framework for automating a major part of the benchmark generation process. Through a case study encompassing three domains (i.e., two single-task domains and one multi-task domain), we demonstrate how to customize the LLM prompts (e.g., omitting system utterances or speaker labels) for those three domains and characterize the generated task-oriented queries. The generated ToQB dataset is made available to the public. We further discuss new domains that can be added to ToQB by community contributors and its practical applications.*

## 1. Introduction

The capabilities of state-of-the-art search engines, virtual assistants, and chatbots are rapidly evolving to help enable seamless task completion for users. As the capabilities of NLP (natural language processing) and GenAI (generative artificial intelligence) expand, there is a growing demand for fulfilling task-oriented queries (ToQs), namely, *action queries*. The rapid expansion from purely informational searches underscores the importance of systems that can execute actions on behalf of the user [18]. Users expect to effortlessly send emails and messages, manage alarms and timers, create notes and lists, control media playback, and even manage their physical environments by controlling lights or locking/unlocking doors [22]. Consequently, the accurate and efficient fulfillments of task-oriented queries have become an essential benchmark for satisfactory, efficient, or delightful human-computer interactions in the era of large language models (LLMs) [8][16] and GenAI.

The recent development of natural language understanding (NLU) and fulfillment systems for task-oriented queries presents a formidable challenge. Such systems must seamlessly integrate with a multitude of external services and applications (e.g., email, messaging, clock, media player, and home automation) to execute user commands. However, a quality concern can arise in the form of hallucinations generated by LLMs [9][20][21][23][24]. Those hallucinations transcend mere quality issues, posing potential safety and privacy risks. An incorrectly processed query could inadvertently open doors (i.e., compromising the user safety) or send sensitive emails to unintended recipients (i.e., harming the user privacy).

Evaluating and optimizing the quality of NLU and fulfillment systems designed for task-oriented queries mandates the use of a dedicated benchmark. Such a benchmark would facilitate iterative improvements and comparisons across various implementations. Unfortunately, the public domain presently lacks a suitable benchmark. While task-oriented dialogue datasets (e.g., TaskMaster [19] and MultiWOZ [12][14]) exist, they capture full conversations, including responses of human agents (or a virtual assistant). It reflects a piecemeal request model where users explain the details of their requests one-by-one (e.g., destination for a taxi, pick-up location, and then pick-up time). In the target scenario of this paper, users however submit one-shot action queries via typing or voice (see Table 1 for a comparison example). The absence of such one-shot action queries benchmark creates a significant gap, hindering speedy research progress and open collaborations in this crucial area.

This paper addresses the challenge by introducing an innovative, automated approach to generating such a benchmark. The presented framework leverages existing task-oriented dialogue datasets as input. It employs an LLM to extract and transform the original user intents into one-shot queries. From the NLP theoretical perspective, the benchmark generation process is equivalent to a novel NLP sub-task for dialogue user request summarization. In the target domain, task-oriented dialogues from the existing NLP benchmarks are fed into an LLM service that is asked to generate one-paragraph summaries of the user requests within dialogues. This automatically generates a large number of queries that summarize original user requests (or intents) in given task-oriented dialogues. An example one-paragraph summary generated by the LLM is showcased in Table 1, along with the input dialogue. Subsequent analysis of each LLM request-response pair (e.g., verification using the annotations of the input dialogues) identifies and filters any responses with interpretation or generation errors.

**Table 1. Examples of a Task-oriented Dialogue and a Task-oriented Query.**

| **Task-oriented Dialogue** (Source: MultiWOZ) |
|---|
| User: Hi, could you help me find a taxi out of **Cocum**? |
| System: Could you tell me when you would like to leave by and when you would like to arrive? |
| User: I need to be picked up by 15:00. Could you provide me the **car type** and **contact number**, please? |
| System: What is your final destination? |
| User: Oh, I'm sorry. I would like to travel to **Gallery at Twelve High Street**. I actually need to be there **by 15:00**. |
| System: OK, I can make that change. |
| User: **When** will the car arrive? |
| System: Unfortunately, it does not give me a specific arrival time. But if you like, you can call them to get that information. Their phone number is 07916703661. |
| User: Thanks, I will **call** them. Can you tell me what type of car I will have? |
| System: Your car type is a white Tesla. |
| **Task-oriented Query** (Generated in this study) |
| Hi, I am looking for a taxi to get me from Cocum to Gallery at Twelve High Street by 15:00. I need to know the car type and contact number, as well as the estimated time of arrival. |

To demonstrate the efficacy of the framework, we conduct a case study using the TaskMaster v2 and MultiWOZ datasets. Through this case study, a high quality benchmark containing 2,922 action queries is successfully generated in an automated fashion, without having to directly crowdsource the raw conversation data. Specifically, the case studies use the 2,922 task-oriented dialogues from the three domains[1] – food ordering, taxi reservation, and multi-tasks (including taxi reservation) – as the input data.

We devise and evaluate three LLM prompt styles. The result reveals that omitting all system utterances in a given dialogue consistently yields a higher success ratio with the used LLM. On the other hand, omitting speaker labels demonstrates a success ratio dependent on the complexity and characteristics of an examined domain. Thus, the user speaker labels are omitted in only two out of the three domains.

The rest of this paper is organized as follows. Section 2 presents our approach of generating action queries from the existing task-oriented dialogues and our framework behind it. Section 3 presents our techniques to automate the generation process using an LLM service. Section 4 conducts a case study by using two existing task-oriented dialogues datasets for three target domains. Section 5 characterizes the generated action queries via lexical, syntactic and semantic analyses. Section 6 discusses the implications, how to extend ToQS, and its applications. Section 7 reviews the related works before concluding the paper in Section 8.

## 2. Approach

This section presents a new NLP (Natural Language Processing) sub-task for emerging action-centric applications and our automated benchmark generation methodology built on top of the new NLP sub-task.

### 2.1. Formulation of New NLP Sub-Task

This subsection defines a new NLP task, namely *dialogue user request summarization*. We note that this target NLP task solely focuses on summarizing the request of a speaker (i.e., the user) within a given dialogue.

Figure 1 shows where this new NLP belongs among other relevant NLP tasks. It is different from typical abstractive text summarization NLP task [13] where the utterances of all speakers would be summarized if the same dialogue data is given as part of an LLM (Large Language Model) input. It is still a sub-task of text summarization task and instruction-based summarization task, where processing the given instruction relies on a deep understanding of conversational contexts and flows (e.g., to extract the intent of a speaker).

**Input and Output Data Examples.** Table 1 shows a pair of the input dialogue and output summary for the call taxi domain in the target NLP task. The associated summarization task request (not shown in the table) explicitly asks a human rater (or an LLM service) to summarize only the user request of the given input dialogue. The output summary correctly summarizes the important elements of the user request: the pickup location ("Cocum"), drop off location ("Gallery at Twelve High Street"), and desired drop off time ("by 15:00"). It also correctly summarizes the interests of the user in getting the car type and contact number information.

---

[1] A domain maps well to a dialogue act in the NLP taxonomy. Domains are also referred to as verticals or journeys in applications or production services using NLP.

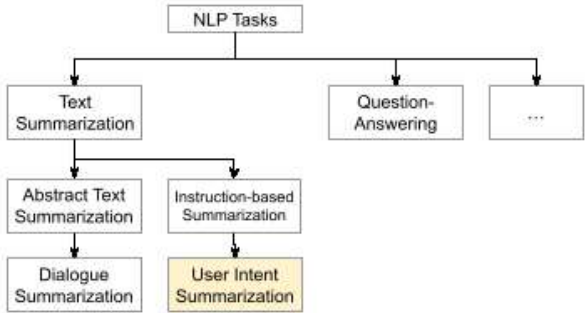

**Fig. 1.** Classification of Relevant NLP Tasks.

The summarized desire for getting the estimated arrival time information, however, is not entirely accurate. While the user initially inquired about estimated arrival time (as specified in the output summary), the system did not have it. Thus, the system instead provided the phone number (in the 4th system utterance in Table 1), making the user want to use the phone number to contact them (the 5th user utterance). Based on that, a more precise summary capturing the conditional aspect would be "I want to know the estimated arrival time, but if unavailable, I will call if a contact phone number can be provided."

The generated summary is supposed to capture the nuanced conditional aspects of the user request but not the system responses themselves (e.g., the situational contexts or environments at a specific moment when the given input dialogue is recorded). This illustrates the high complexity inherent in this new NLP task. That is, it would be a time-consuming, error-prone task if such user intent summarization tasks are done manually by humans.

### 2.2. Automated Generation Methodology

This subsection presents the automatic generation methodology of a task-oriented queries benchmark using an LLM service. Our methodology leverages existing task-oriented dialogue datasets as input. It employs an LLM to extract and transform the original user intents into one-shot action queries.

Specifically, task-oriented dialogues from the existing NLP benchmarks are fed into an LLM service that is asked to generate one-paragraph summaries of the user requests within dialogues. This automatically generates a large number of action queries that summarize original user requests (or intents) in task-oriented dialogues.

For each summarization task, the employed LLM service generates a one-paragraph summary. We verify each LLM request-response pair and filter response candidates with hallucinations by using tools developed specifically for this purpose. Then subsequent analysis further characterizes the generated queries.

### 2.3. Framework

The overall process of the presented methodology, illustrated in Figure 2, consists of the following five key steps:

**Step 1. Read Annotated Dialogues.** It uses a task-driven dialogues dataset as an input where each input dialogue represents a request for one or multiple types of tasks. Each dialogue consists of a sequence of transcribed user and system utterances. Optionally, it contains annotations for the domain and key slots. Here, domain is equivalent to dialogue act in the NLP taxonomy. During this step, our framework parses files in the given input dataset to read each dialogue in the dataset.

**Step 2. Preprocess Dialogues and LLM Prompt Engineering.** For each dialogue, the framework preprocesses all the read utterance data and generates an LLM request. It uses the following LLM prompt template in order to convert the preprocessed utterance data into a format suitable for the LLM requests:

> In the following [DIALOGUE], the USER has a conversation with SYSTEM. Pretend you're the USER. Summarize and say the request of the USER in 1 paragraph.
>
> [Dialogue]:
>
> ... (Preprocessed utterance data) ...

This demonstrates that the same template can be applied across multiple domains. Therefore, our approach is inherently horizontal and easily scalable to various domains.

**Step 3. Summarize using LLM.** For each generated LLM request, the framework calls an LLM API (Application Programming Interface) service in order to extract and summarize the user request based on the preprocessed utterance data in the LLM request. The framework takes response candidates from the LLM service or raises an error if no valid LLM response is received. The result is then stored in a local file system for the next step.

**Step 4. Check and Classify LLM Responses.** This step involves automatic verifications and manual reviews of the captured LLM response candidates to assess the success of the summarization and identify any contained hallucinations. The automatic verifications are conducted if and only if the annotations extracted from the input dialogue data are available. Using the annotations, automatic verifications, for example, check whether the key slots are correctly summarized in given responses [31]. The manual reviewers can also utilize the annotations to expedite the cross-checking and classification processes. The manual reviewers are responsible for identifying hallucinations for unsuccessful responses.

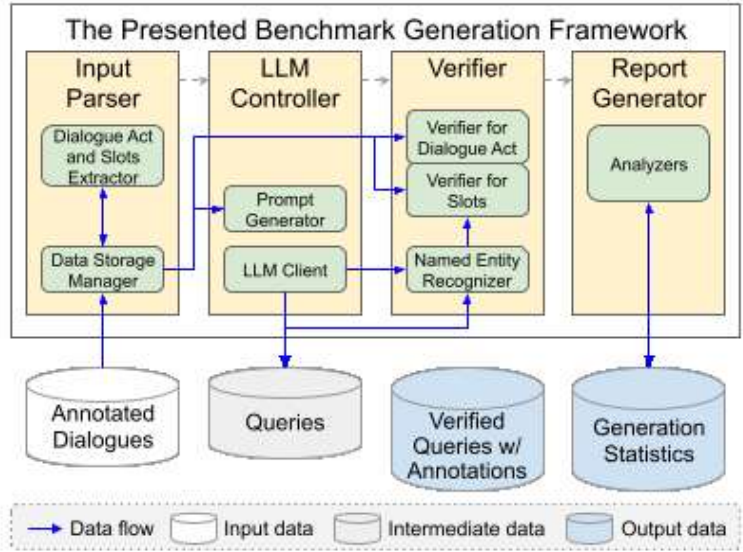

**Fig. 2. Overview of the presented automated generation process of task-oriented queries benchmark.**

**Step 5. Analyze Statistics.** Once all the input dialogues are processed, the framework automatically analyzes the generated queries data using multiple scripts. For example, scripts are developed and used to help us conduct lexical, syntactic and semantic analyses.

## 3. Techniques

This section presents the three key techniques designed and used to ensure the quality of the generated queries dataset.

**Preprocessing Options.** Let us explore the following three options for preprocessing input dialogue utterance data:

- **User & System Option (U&S)** to explicitly specify both the user and system utterances in a respective LLM request. It is typically done by formatting each dialogue as "User: (utterance) \newline System: (utterance) ...".

- **User Option (User)** to specify only the user utterances, excluding the system utterances. The format involves listing the user utterances as "User: (utterance) \newline User: (utterance) ..." in a respective LLM request.

- **User w/o Speaker Option (UwoS)** to specify only the user utterances after eliminating the speaker labels altogether. As a result, it is typically in the form of "(user utterance). (user utterance). ...". Any sentences of the original user utterances are appended with a default punctuation mark (i.e., ".") if those sentences lack any trailing punctuation marks. All sentences in the user utterances are thus concatenated into a single paragraph and embedded in a respective LLM request.

The first two options highlight a key tradeoff. Including the system utterances in a generated LLM request may lead to an LLM service generating a summary with extraneous information beyond the exact user request. Conversely, omitting the system utterance can potentially hinder the ability of a used LLM service to fully grasp the user intents conditionally expressed upon specific system responses.

**Annotation Cleansing.** The annotations in the input dialogue datasets require manual data cleansing mainly due to:

1. Transcription or typing errors
2. Extra words (e.g., some)
3. Synonyms or abbreviations (e.g., peas for beans, coke for pepsi, and bbq for barbeque)
4. Grammar or inconsistency issues (e.g., "couple of can cokes" vs. "couple of cans of coke")
5. Approximated phases for the same meaning in a same dialogue (e.g., "two hash browns" vs. "some hash brown potatoes")

Table 2. Characteristics of Input Dialogue Datasets.

| Task ID | Domain | Number of Dialogues | Average number of turns per dialogue ± standard deviations |
|---|---|---|---|
| *Food* | Order Food | 1,050 | 13.3 ± 4.21 |
| *Taxi(S)* | Call Taxi (Single task) | 430 | 7.7 ± 2.07 |
| *Taxi(M)* | Call Taxi (Multitasks) | 1,442 | 19.0 ± 4.26 |

6. Long sentences intended to represent itemized slots (e.g., "one order of shrimp i mean of chicken pasta")
7. Unintended information derived from the system utterances (i.e., not part of the user request)

By addressing these issues through an annotation cleansing process, this study aims to ensure greater consistency and accuracy in annotating user requests. Specifically, while LLMs are usually tolerant of minor errors in input datasets, the verification step is not (e.g., when using text matching), unless verification operations also utilize LLMs or equivalent techniques (e.g., those that understand synonyms and different expressions of the same concept).

**LLM Configurations.** In this study, all the experiments use the LLM service that operates in a cloud data center. It is powered by a publically accessible LLM with 35 billion parameters.

The used LLM service can generate up to 8 response candidates per request. Each response candidate has an associated ranking score. The framework takes the highest score response candidates that pass all the verifications and stores them for the follow-up manual analyses. Usually the highest score response candidate that passes the automated verifications is taken.

If no such a candidate exists, the framework then retries the LLM-based generation flow up to 10 times. In practice, by using 3 retires, we are able to generate action queries almost all the used input dialogues.

## 4. Case Study

To enable robust quantitative analysis, this study leverages the two publically available, task-oriented dialogue datasets with a significant number of dialogues (e.g., >1,000). An alternative is to manually generate dialogues data (e.g., via crowdsourcing). However, it is costly especially due to the additional manual review required for formatting, checking, correcting, and annotating the manually generated dialogues.

The input dialogue datasets are carefully chosen from MultiWOZ 2.2 [12] and TaskMaster v2[2]. Those datasets are selected because they have fully annotated, task-driven dialogues. Table 2 outlines the key features of the selected input dialogue datasets.

### 4.1. Domains

Within TaskMaster, the chosen domain is food ordering (namely, *Food*), while the other domains (e.g., music playing, restaurants, movies, and sports) are excluded because those domains are by users with search intents[3]. All 1,050 dialogues from the selected Food domain are evaluated. We note all dialogues within TaskMaster are single-task focused.

Within MultiWOZ, the chosen domain is Taxi (i.e., taxi reservation). Here, some dialogues are for multiple tasks, including Taxi. All 1,872 dialogues from the selected Taxi domain are examined where 430 dialogues are for single task (namely, *Taxi(S)*) and the remaining 1,442 dialogues are for multiple tasks (namely, *Taxi(M)*). For example, a dialogue for the Taxi(M) domain can involve a user requesting actions like hotel and restaurant reservations in addition to booking a taxi between those two places. In total, the chosen datasets consists of 2,922 dialogues across three domains: Food, Taxi(S), and Taxi(M).

### 4.2. Prompt Engineering

Before collecting the summarized action queries, let us optimize how the LLM service is invoked for the target NLP task through prompt engineering, i.e., tailoring prompts to elicit the desired LLM responses, focusing on summarizing the original user requests:

**Omit System Utterances.** To identify the optimal approach for utilizing user and system utterances in the target domain, we first compare the success ratios of the three utterance preprocessing options: User & System (U&S), User, and User w/o Speaker (UwoS). We note that U&S incorpo-rates both user and system utterances, while User and UwoS only utilize user utterances.

---

[2] Available at
https://research.google/resources/datasets/taskmaster-2/

[3] Dialogues with search intents typically result in search queries that can still be effectively fulfilled by existing search or information retrieval (IR) engines. Since search queries and their benchmarks are extensively studied in the IR field, this paper focuses on generating non-search, action-oriented queries.

The U&S option exhibits the lowest success ratios across all three domains. The two other preprocessing options, i.e., User and UwoS, consistently demonstrate the higher success ratios than U&S for all three domains. Here, the confidence intervals estimate population proportions at the 95% confidence level. Indeed, the U&S option proves ineffective mainly due to its high probability of generating extraneous information that is typically derived from the system utterances specified in a respective LLM request.

**When to Label Speaker(s)?** We delve into comparing the remaining two preprocessing options: User and UwoS. By analyzing their respective success ratios, we aim to better understand when and why one option outperforms the other in achieving a higher success ratio. We note that the User option retains the user speaker labels in a derived LLM request, while the UwoS option omits them entirely.

For the Food and Taxi(S) domains, the UwoS option demonstrates higher success ratios than the User option. However, the scenario shifts when we examine the most complex domain, Taxi(M). Here, the User option prevails with the higher success ratio than that of the UwoS option.

## 5. Characterization

This section characterizes the generated task-oriented queries through lexical, syntax, and semantic analyses.

### 5.1. Lexical Analysis

The lexical analysis counts the characters, words, and sentences in the generated queries. Their cumulative distribution functions (CDFs) and probability density functions (PDFs) are plotted in Figure 3 and 4, respectively. In all cases, the Taxi(S) domain has the shortest outputs, while Taxi(M) has the longest outputs (e.g., see the peaks and tails in Figure 4). The result is aligned with the number of dialogues in the inputs (shown in Table 2), i.e., the output query length is proportional to the number of dialogues in the input.

The rest of this subsection further analyzes the distributions in details so as to identify major thresholds. Those thresholds (e.g., 50 percentile) could be used to choose a subset of the generated action queries for some specific user cases. For example, if short succinct (or long descriptive) queries are required, then one can select a certain number of such queries by using the percentile data.

**The number of characters including spaces.** Figure 3(a) and 4(a) depict the CDF and PDF, respectively, of character counts of the generated queries. Here, the x-axis represents the number of characters in each query. The distribution reveals the peak at around 125 characters for Taxi(S), around 165 characters for Food, and around 310 characters for Taxi(M).

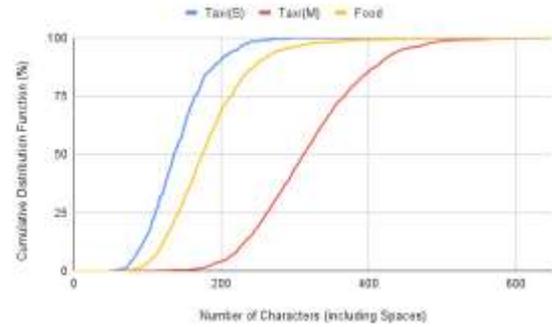

(a) CDF for the number of characters

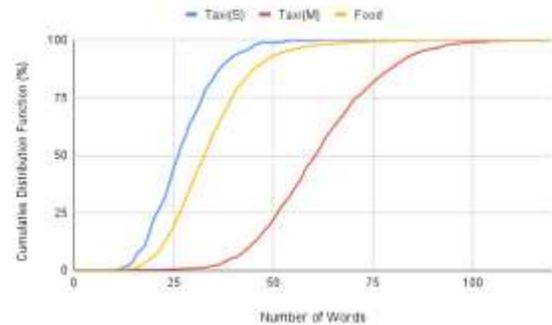

(b) CDF for the number of words

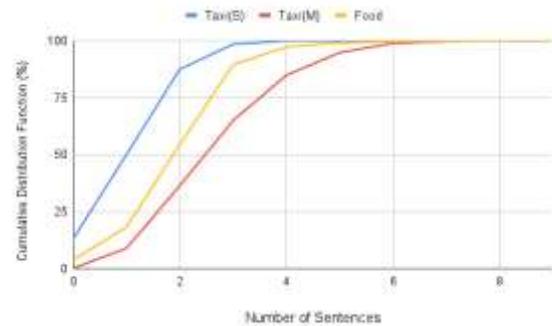

(c) CDF for the number of sentences

**Fig. 3.** Cumulative distribution functions (CDFs) for lexical analysis.

- The 10 percentile is 91 characters, 118 characters, and 226 characters for the Taxi(S), Food, and Taxi(M) domains, respectively. Some example queries are: *"I need to book a taxi from Hakka to La Margherita by 13:45."* (59 characters) for Taxi(S); *"I would like to order two poke aloha bowls and four poke tacos for take-out. Thank you."* (87 characters) for Food; and *"I am looking for a restaurant called Thanh Binh that can accommodate 6 people on Friday at 12:00. I would* *also like to book a taxi for 6 people from the concert hall to the restaurant at 11:00."* (193 characters) for Taxi(M).

- Similarly, the 25 percentile is 112 characters, 143 characters, and 263 characters for Taxi(S), Food, and Taxi(M), respectively.

- The median is 137 characters, 175 characters, and 312 characters for Taxi(S), Food, and Taxi(M), respectively.

- The 75 percentile is 165 characters, 214 characters, and 368 characters for Taxi(S), Food, and Taxi(M), respectively.

- The 90 percentile is 198 characters, 252 characters, and 419 characters for Taxi(S), Food, and Taxi(M), respectively. Some example queries are: *"I am requesting a taxi from Finches Bed and Breakfast to Don Pasquale Pizzeria. I would like to arrive at 1pm to meet my husband for lunch. I would like the contact information for the driver so that I can reach them if necessary. Thank you."* (241 characters) for Taxi(S); *"I would like to place an order for Indian food for three people. I would like to order beef shish kebabs for one person, chicken tandoori for one person, and green curry with chicken for the third person. I would also like to add garlic to all of the dishes."* for Food; and *"I am planning a trip to Cambridge and I need your help with booking a restaurant and a taxi. I am looking for an expensive restaurant serving British food in the west area of Cambridge. I would like to book a table for one on Tuesday at 7:30 PM. I am also looking for an attraction in the same area. Can you recommend one and provide me their phone number? I would like to take a taxi between the restaurant and the attraction. I would like to leave the attraction by 7:15 PM."* (476 characters) for Taxi(M).

**The number of words.** Figure 3(b) and 4(b) depict the CDF and PDF, respectively, of word counts of the generated queries, where the x-axis represents the number of words in each query. The distribution reveals the peak at around 25 words for Taxi(S), around 30 words for Food, and around 55 words for Taxi(M).

- The 10 percentile is 18 words, 23 words, and 44 words for the Taxi(S), Food, and Taxi(M) domains, respectively. Some example queries are: *"I am requesting a taxi to Tandoori Palace. I would like to leave after 16:30. Thank you."* (17 words) for Taxi(S); *"I would like to order chicken fettuccine Alfredo, garlic bread, and a 2-liter of Mountain Dew. I am ready to pay. Thank you."* (23 words) for Food; and *"I am looking for a free college and a cheap Mediterranean restaurant in Centre. I would like to know the address, postcode, phone number, and contact information for a taxi company that can take me from the college to the restaurant by 14:45."* (43 words) for Taxi(M).

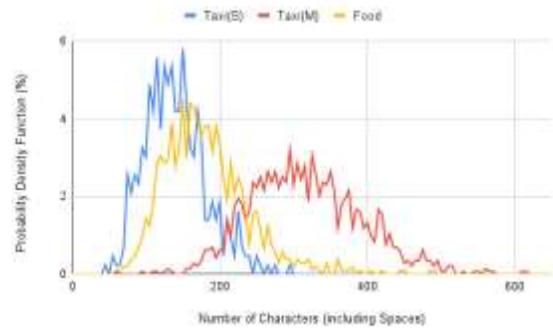

(a) PDF for the number of characters

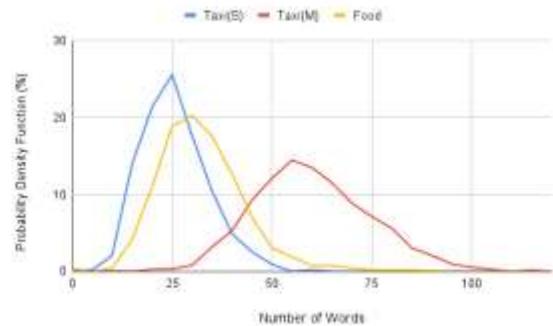

(b) PDF for the number of words

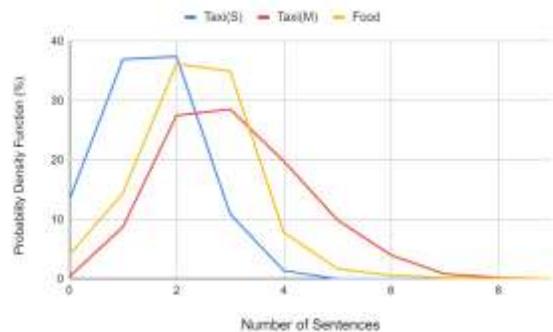

(c) PDF for the number of sentences

**Fig. 4.** Probability density functions (PDFs) for lexical analysis.

- Similarly, the 25 percentile is 21 words, 27 words, and 52 characters for Taxi(S), Food, and Taxi(M), respectively.

- The median is 27 words, 33 characters, and 61 characters for Taxi(S), Food, and Taxi(M), respectively.

- The 75 percentile is 33 words, 40 words, and 71 words for Taxi(S), Food, and Taxi(M), respectively.

- The 90 percentile is 38 words, 48 words, and 82 words for Taxi(S), Food, and Taxi(M), respectively. Some example queries are: *"I would like to book a taxi from Saint Johns Chop House to Williams Art and Antiques. The taxi should leave after 3:15 PM. I apologize for the error in my previous request. Thank you for your assistance."* (38 words) for Taxi(S); *"I would like to place an order for Indian food for three people. I would like to order beef shish kebabs for one person, chicken tandoori for one person, and green curry with chicken for the third person. I would also like to add garlic to all of the dishes."* (50 words) for Food; and *"I am planning a trip to Cambridge and I need your help with booking a restaurant and a taxi. I am looking for an expensive restaurant serving British food in the west area of Cambridge. I would like to book a table for one on Tuesday at 7:30 PM. I am also looking for an attraction in the same area. Can you recommend one and provide me their phone number? I would like to take a taxi between the restaurant and the attraction. I would like to leave the attraction by 7:15 PM."* (93 words) for Taxi(M).

**The number of sentences.** Figure 3(c) and 4(c) depict the CDF and PDF, respectively, of sentence counts of the generated queries, where the x-axis represents the number of sentences in each query. The distribution reveals the peak at around 2 sentences for Taxi(S), around 2 sentences for Food, and around 3 sentences for Taxi(M).

- The median is 1 sentence, 2 sentences, and 3 sentences for the Taxi(S), Food, and Taxi(M) domains, respectively. Some example queries are: *"I would like to book a taxi from Jesus College to Norwich train station after 2:30 PM today."* (1 sentence) for Taxi(S); *"I would like to order takeout food for three people. I would like one Big Mac, one chili burger, one patty melt, three large french fries, and three cans of Mountain Dew Code Red."* (2 sentences) for Food; and *"I am looking for a 3-star hotel in the expensive price range for 7 people for 5 nights beginning on Wednesday. I would like to book the Lensfield hotel. I also need information on Saint Johns College and a taxi to commute between the two places at 11:45."* (3 sentences) for Taxi(M).

- Similarly, the 75 percentile is 2 sentences, 3 sentences, and 4 sentences for Taxi(S), Food, and Taxi(M), respectively.

- The 90 percentile is 3 sentences, 4 sentences, and 5 sentences for Taxi(S), Food, and Taxi(M), respectively. Some example queries are: *"I need a taxi from Peking Restaurant to Maharajah Tandoori Restaurant. I need to arrive at 10:45. I would prefer a blue Audi."* (3 sentences) for Taxi(S); *"I would like to order two gyros and one Greek salad for delivery. The gyros should be for two people. I would also like to add on one of those Greek salads. Thank you."* (4 sentences) for Food; and *"Hi, I am looking for an upscale, expensive restaurant in the centre of town that serves British food. I need a reservation for 8 people on Monday at 11:45. I also need a 3-star hotel in the same price range with wifi. I need a reservation for 8 people staying 5 nights starting Monday. Please book a taxi from the hotel to the restaurant by 11:45. I need the contact number and car type. Thank you for your help."* (7 sentences) for Taxi(M).

### 5.2. Syntax Analysis

The syntax analysis identifies common patterns in the first and last sentences of generated queries.

**The styles of first sentences.** The most commonly observed styles of the first sentences of the generated queries are as follows:

- "Hi, …"
- "I am requesting …"
- "I would like to [book|request|order] …"
- "I need [to [book]] …"
- "I'm calling to [place an order|order] …"
- "I am [placing|ordering] …"
- "I am looking for …"
- "I am [planning|visiting] …"

Here, [<word 1>] means <word 1> is optional. It can be nested, e.g., formulated as [<word 1> [<word 2>]]. Similarly, [<word 1> | <word 2>] means either <word 1> or <word 2> can be presented.

**The styles of last sentences.** The most commonly observed styles of the last sentences of the generated queries are as follows:

- "… Thank you [.|!|very much!|for your [assistance|help|time]]"

**Table 3.** Most frequent words (excluding Arabic numbers) sorted by their occurrence frequencies in the generated task-oriented queries of each of the three target domains.

| Food | | | Taxi(S) | | | Taxi(M) | | |
|---|---|---|---|---|---|---|---|---|
| 1st–35th | 36th–70th | 71st–105th | 1st–35th | 36th–70th | 71st–105th | 1st–35th | 36th–70th | 71st–105th |
| I | *Salad* | From | To | *Restaurant* | *Assistance* | The | *An* | *Place* |
| **Like** | Not | *Dogs* | I | Type | *La* | I | *Phone* | *Cheap* |
| **Would** | On | *Bread* | **Taxi** | College | **Can** | To | *Parking* | *West* |
| To | **Be** | *Ketchup* | The | Pm | *Guest* | A | *Postcode* | *Contact* |
| **Order** | *Fries* | Else | From | **Know** | Two | For | *Area* | *Range* |
| And | *Orders* | Anything | **Would** | Please | *Curry* | **Would** | *Guesthouse* | *Saturday* |
| For | *Your* | **Take-out** | A | *Station* | *Noddle* | **Like** | *Reservation* | *Places* |
| A | *I'm* | Cokes | **Like** | *Cambridge* | Other | And | *Museum* | Be |
| Of | **Need** | *Sandwich* | And | Not | **Requests** | In | Between | *Priced* |
| One | **Have** | *Breakfast* | Me | **Request** | So | *Restaurant* | *Table* | Car |
| Two | *Pizza* | Please | **Need** | **Do** | **Appreciate** | At | **Take** | **Provide** |
| *You* | *Sauce* | *Thai* | By | **Go** | *Bistro* | Am | Please | *South* |
| **Thank** | *Rice* | *Pork* | At | **Will** | **Get** | Also | *Wifi* | *Thursday* |
| With | **Do** | *Diet* | Am | Time | If | *Taxi* | **Get** | Two |
| The | **Calling** | *Dog* | You | Any | *Jesus* | Hotel | College | **Visit** |
| **Takeout** | **Looking** | Any | **Thank** | With | *Park* | Of | *Expensive* | *Wednesday* |
| Three | *Soup* | Last | **Requesting** | **Have** | *Fen* | **Looking** | *Finally* | *House* |
| *Food* | *Burrito* | *Onions* | **Arrive** | **Picked** | It | **Book** | *Leave* | *Sunday* |
| Am | It | *Lettuce* | For | **Provide** | **Looking** | **Need** | Attraction | **Make** |
| *People* | **Help** | *Tomato* | Up | *Hotel* | **Should** | On | **Can** | *City* |
| Also | *Burritos* | Medium | **Book** | *Train* | There | From | Same | *East* |
| Large | **Ordering** | *Bacon* | After | *Gallery* | Also | *People* | If | *Centre* |
| *Chicken* | **Place** | *Beans* | Contact | *Bar* | *Company* | Me | **Find** | *Moderately* |
| *Side* | Extra | *Egg* | Number | That | *Driver* | Number | *Time* | *Friday* |
| *Is* | *Coke* | *Gyros* | **Take** | *City* | *Green* | By | *Is* | *Tuesday* |
| *That* | In | Same | Of | *Museum* | *Junction* | With | *Type* | *Monday* |
| My | **Will** | Should | *Alpha-milton* | *Saint* | *Kings* | *Town* | *You* | **Commute** |
| An | **Take** | *Sour* | Your | *Bed* | *Let* | *Free* | *Fee* | *Information* |
| *Cheese* | *Barbecue* | *Dressing* | House | *Breakfast* | *Pool* | *Nights* | *Entrance* | **Will** |
| Up | *Beef* | *Sausage* | **Be** | **Depart** | *Street* | **Know** | *Star* | *Food* |
| Hi | Out | Minutes | **Pick** | *Lodge* | *Tandoori* | *Center* | Go | *Stay* |
| **Pick** | *Mustard* | *Tomatoes* | **Leave** | *Church* | *Theatre* | *Cambridge* | *Price* | **Arrive** |
| Hot | Some | *French* | Help | **Is** | *Today* | That | *There* | My |
| *Person* | **Add** | *Bowl* | Additional | My | *Avalon* | **Starting** | It | As |
| All | **Placing** | *Fried* | Car | As | *Brasserie* | *Address* | *North* | *Room* |

All the common styles are for the thank you sentences.

### 5.3. Semantic Analysis

The semantic analysis identifies the most common words (e.g., verbs and nouns) in the generated queries. As highlighted in bold texts in Table 3, the top-10 most common verbs are:

- For the Taxi(S) domain: "would", "like", "need", "am", "thank", "requesting", "arrive", "book", "take", "be"
- For the Taxi(M) domain: "would", "like", "looking", "book", "need", "free", "know", "starting", "take", "get"
- For the Food domain: "like", "would", "order", "thank", "takeout", "pick", "be", "need", "have", "do"

Here, the following verbs are seen in the top-10 verb lists of all three domains: "would", "like", and "need". Those three verbs are used in the common styles of first sentences.

The top 10 most common nouns are:

- For Taxi(S) domain: "Taxi", "Contact", "Number", "Alpha-milton", "Car", "Restaurant", "College", "Station", "Cambridge", "Hotel"

- For the Taxi(Multi) domain: "Restaurant", "Taxi", "Hotel", "People", "Me", "Number", "Town", "Nights", "Center", "Cambridge"
- For the Food domain: "You", "Food", "People", "Chicken", "Side", "Cheese", "Person", "Salad", "Fries", "Orders"

Unlike the common verbs, there is no common noun across the top 10 noun lists of the three domains because nouns are highly context-dependent.

One can use the verbs in Table 3 to infer common action types modeled in ToQB, and the nouns to understand the context and slots associated with the queries in the benchmark. This information could help ToQB users decide when, where, and how to use specific queries in each domain. For example, one can choose queries containing the word "breakfast" to evaluate food ordering queries in the morning. Similarly, one can choose queries containing the word "Cambridge" to model queries from a specific geographical location.

## 6. Discussions

This section discusses the availability and anticipated applications of the presented ToQB (Task-oriented Queries Benchmark).

**Benchmark Availability.** The ToQB datasets will be made available to the public at:

https://github.com/google/task-oriented-queries

In addition to a large number of the generated action queries, one extra benefit of ToQB is that users can find the annotations (e.g., dialogue act and slots) from the other used task-oriented dialogue datasets. We welcome contributions from both the research community and industry to incorporate into future versions of ToQB.

**Call for Contributions.** The automated benchmark generation methodology and framework presented in this paper can be readily adapted to create task-oriented query datasets for a wide array of domains. By utilizing existing (or newly collected) task-oriented dialogue datasets as input and customizing the LLM prompts as demonstrated in our case studies, researchers and practitioners can generate high-quality, action queries benchmarks tailored to specific domains. Some example domains that would be of particular interest to the community include:

- **Home Automation.** Generating queries for controlling home automation devices (lights, thermostats, door locks), setting up routines and schedules, monitoring energy usage, troubleshooting connectivity issues, and integrating new devices into the system.
- **News.** Generating queries for searching for news articles, filtering by topic or source, reading summaries, getting personalized recommendations, setting news alerts, saving articles for later, and finding related content.
- **Multimedia.** Creating benchmarks for tasks like finding movies or TV shows, getting showtimes, purchasing tickets, discovering new music, creating playlists, listening to podcasts, and searching for events or concerts.
- **Productivity.** Generating queries for setting reminders, creating to-do lists, managing calendars, scheduling meetings, organizing emails, taking notes, tracking projects, and collaborating with others.
- **Geo.** Generating queries for finding directions, searching for nearby points of interest (restaurants, gas stations, etc.), estimating travel times, getting traffic updates, finding parking, and exploring public transportation options.
- **Health and Fitness.** Creating benchmarks for tasks like tracking workouts, monitoring fitness goals, logging food intake, finding healthy recipes, setting reminders for medication or appointments, and accessing health resources.
- **Social Media.** Generating queries for posting updates, searching for friends or groups, sending messages, commenting on posts, sharing content, managing privacy settings, and exploring trending topics.
- **Weather.** Generating queries for checking current weather conditions, getting forecasts, setting up weather alerts, finding historical weather data, planning outdoor activities based on weather, and understanding weather-related terminology.
- **Sports.** Creating benchmarks for tasks like getting scores and schedules, following favorite teams or athletes, purchasing tickets, finding sports news or analysis, researching statistics, and participating in fantasy sports leagues.
- **Events and Ticketing.** Creating benchmarks for tasks like searching for events (e.g., concerts, festivals, and conferences), purchasing tickets, checking seating availability, finding transportation or accommodation options, and managing event registrations.

The above list highlights areas where the existing virtual assistants and chatbots offer some level of support. The following list focuses on anticipated future domains that will become feasible as LLM capabilities and infrastructure mature:

- **Healthcare.** Generating queries for appointment scheduling, medication refills, symptom inquiries, and general health information requests.
- **Finance.** Creating benchmarks for tasks like checking account balances, transferring funds, paying bills, tracking expenses, monitoring investments, and inquiring about final products and investment options.
- **Travel.** Generating queries for booking flights, hotels, and rental cars, as well as inquiring about itineraries, destinations, and local attractions.
- **Customer Service.** Developing benchmarks for common customer service interactions, such as tracking orders, requesting refunds, or troubleshooting product issues.
- **Shopping and E-commerce.** Generating queries for browsing products, searching for specific items, comparing prices and features, adding items to cart, completing purchases, managing orders, tracking shipments, and requesting customer support.
- **Education.** Creating benchmarks for tasks like enrolling in courses, accessing learning materials, submitting assignments, communicating with instructors, seeking academic advising, and exploring educational resources.
- **Job Search and Career Development.** Creating benchmarks for tasks like searching for job postings, applying for jobs, building resumes, preparing for interviews, researching companies, networking with professionals, and exploring career paths.
- **Government and Civic Services.** Generating queries for accessing government information and services, finding local representatives, registering to vote, paying taxes, applying for licenses or permits, and reporting issues or concerns.
- **Legal Services.** Generating queries for finding legal information, searching for lawyers or law firms, scheduling consultations, inquiring about legal processes or documents, and seeking legal advice on specific topics.
- **Real Estate.** Creating benchmarks for tasks like searching for properties to buy or rent, filtering by criteria (e.g., price, location, and features), scheduling viewings, contacting real estate agents, calculating mortgage payments, and researching property values.
- **Gaming.** Generating queries for finding games or platforms, searching for walkthroughs or guides, connecting with other players, troubleshooting technical issues, purchasing in-game items, and staying up-to-date on game news and releases.
- **Home Improvement and Repair.** Creating benchmarks for tasks like finding DIY instructions or tutorials, searching for contractors or service providers, requesting quotes, scheduling appointments, troubleshooting household issues, and ordering supplies or materials.

By expanding ToQB to encompass those and other relevant domains, our research community can foster further advancements in natural language understanding and task fulfillment capabilities, leading to more effective and versatile AI-powered services.

**Applications.** ToQB offers a wide range of practical applications in the development and evaluation of various LLM-based services:

- **Voice Assistants.** ToQB can be used to assess and improve the accuracy of voice assistants [26] in understanding and fulfilling task-oriented user queries: both voice and text queries. By training and evaluating models on ToQB, developers of virtual assistants can enhance the ability of voice assistants to handle complex, multi-step tasks accurately.
- **Search Engines.** Web or on-device search engine providers can leverage ToQB to evaluate and optimize their services for understanding and responding to action-oriented queries. This can lead to improved search results and a more seamless user experience when users seek to accomplish tasks directly through search.
- **Chatbots.** ToQB is a valuable resource for chatbot developers, enabling them to assess and enhance the natural language understanding (NLU) and task fulfillment capabilities of their chatbots. By training chatbots on ToQB, developers of chatbots can ensure that their chatbots can effectively handle task-oriented requests, even in diverse and complex conversational contexts.

That is, ToQB is well suited for evaluating certain capabilities of LLMs and other multi-modal models. By adding more domains, ToQB can serve as a standardized benchmark for evaluating the performance of LLMs in under-standing and responding to task-oriented queries. This can guide researchers and developers in identifying strengths and weaknesses of different LLM architectures and training strategies.

Overall, ToQB has the potential to significantly advance the field of task-oriented query processing. By providing a comprehensive and standardized benchmark, it facilitates the development of more accurate, efficient, and user-friendly LLM-based services across a wide range of applications. In essence, ToQB, as an open benchmark platform for public and community collaboration, provides a foundation for reproducible measurements essential for developing reliable and high-quality GenAI software [17] for our society.

# 7. Related Work

This section summarizes the related works that are grouped into:

**Language Model-based Services.** The emergence of language models like BERT [2] and LaMDA [3], built upon the transformer architecture [1], has significantly advanced natural language processing (NLP) capabilities. These models, particularly ones trained on massive datasets like GPT-3 [4], have demonstrated emergent abilities [5] to perform various tasks, including generating coherent and contextually relevant responses [3][4][5]. Recent studies have shown that LLMs can even outperform experts in certain annotation tasks [6]. Additionally, the application of LLMs extends beyond language generation, with models like AlphaCode [7] demonstrating competition-level code generation capabilities.

**Challenges in Quality Evaluation.** However, despite their impressive performance, LLMs are prone to generating factually incorrect or nonsensical outputs, often referred to as hallucinations [8][9]. This issue [33][34] is particularly critical in task-oriented applications, where incorrect responses can have real-world consequences [10]. Several studies have investigated the problem of hallucination in text generation [9][25][28][29][32][44] and neural translation [35], particularly in the context of abstractive summarization [11][27][30].

**Existing Approaches for Quality.** Techniques like retrieval-augmented generation (RAG) [36] and content-matching constraints [38] are proposed to mitigate issues with factual accuracy and quality in generated text. Additionally, research has focused on developing metrics to assess the factual accuracy of generated text [39][40][41][42][43]. This work builds a foundation for such existing quality approaches, as the resulting ToQB datasets can be directly used to evaluate different quality techniques. Moreover, researchers and practitioners can use the presented benchmark generation methodology and framework to generate more tailored queries for their own quality evaluation purposes.

**Task-oriented Dialogues.** Our work contributes to this body of research by focusing specifically on the generation and evaluation of task-oriented queries. While existing task-oriented dialogue datasets like MultiWOZ [12][14] and TaskMaster [19] provide valuable resources, they primarily focus on full dialogues rather than one-shot queries.

**Abstractive text summarization.** The devised user request summarization NLP task bears some resemblance to the existing abstractive summarization tasks [13][37] in a sense that both aim to summarize given texts. The key difference is that while the existing tasks summarize the entire text (e.g., all the utterances of all speakers if the text is about a dialogue), the presented NLP task is supposed to specifically capture the request of a single speaker expressed in the dialogue text.

**TODSum.** The TODSum dataset [15] does address task-oriented dialogue summarization but does not specifically target the extraction of user intents for one-shot query generation. Our approach leverages LLMs to bridge this gap, creating a benchmark specifically tailored for evaluating the performance of LLM-based services in handling task-oriented requests.

TODSum used task-oriented dialogues as input for text summarization tasks and produced the task-oriented dialogue summarization benchmark dataset. There are two key differences between TODSum and the presented ToQ Benchmark. While TODSum focuses summarizing a dialogue, the presented benchmark focuses on summarizing only the original request (or intent) of one speaker (i.e., the user). For example, if the original intent of a user is to order Diet Coke but the user ended up order Pepsi because of the availability or the suggestion by the system, TODSum summarizes that the ordered menu item is Pepsi, while ToQ summarizes the original request is diet coke and the second choice is Pepsi if Diet Coke is not available. It reveals the difference in their potential applications (e.g., TODSum to summarize whole dialogues vs. ToQB to generate action queries for evaluating search engines, virtual assistants, and chatbots). Moreover, TODSum only used the MultiWoZ dataset as input, while the presented ToQ benchmark uses both MultiWoZ and TaskMaster as inputs.

**LLM Alignment.** Existing LLM alignment methods utilize human supervisory signals as feedback or employ AI for supervision, enabling LLMs to better model the preferences and values of a target group of humans in specific situations. Those alignment techniques can be classified into three types based on their goals: instruction alignment, hu-man preference alignment, and value alignment. Notably, instruction alignment methods [45][46][47] can utilize ToQB and other similar task-oriented query datasets.

## 8. Conclusion

This paper presented a novel, automated methodology for generating Task-oriented Queries Benchmark (ToQB). By leveraging existing task-oriented dialogue datasets and harnessing the power of LLMs, we have successfully addressed the gap in benchmark resources for evaluating one-shot action queries. The generated ToQB dataset, containing 2,922 diverse action queries across three distinct domains, is a valuable resource for the NLP community and industries aiming to enhance the quality and safety of LLM-based services.

Through a comprehensive case study, we have not only demonstrated the effectiveness of our framework but also highlighted the importance of careful prompt engineering in extracting accurate user intents. Our findings underscore the nuanced relationship between the inclusion of system utterances and speaker labels in LLM prompts, revealing how these factors can significantly influence the success rate of user request summarization.

The lexical, syntactic, and semantic analyses conducted on the generated queries offer valuable insights into the characteristics of task-oriented language. These insights can further guide the development and refinement of NLU (natural language understanding) and fulfillment systems, particularly in understanding the linguistic patterns associated with user requests.

The public availability of the ToQB dataset is expected to spur innovation and research in the domain of task-oriented query fulfillment. By providing a standardized benchmark, we anticipate advancements in search engines, voice assistants, and chatbots, ultimately leading to more accurate, efficient, and user-friendly experiences. Moreover, the demonstrated methodology paves the way for future extensions of the ToQB, accommodating a wider range of task domains and linguistic variations.

Future work will focus on refining the methodology by incorporating more sophisticated annotation techniques and exploring the use of more advanced LLMs. Additionally, we aim to and encourage others to expand the ToQB to encompass a wider range of tasks and domains, thereby further enriching the landscape of benchmark resources for task-oriented queries.

## References

[1] A. Vaswani, N. Shazeer, N. Parmar, J. Uszkoreit, L. Jones, A. N. Gomez, L. u. Kaiser, and I. Polosukhin, "Attention is all you need," in *Advances in Neural Information Processing Systems*, I. Guyon, U. V. Luxburg, S. Bengio, H. Wallach, R. Fergus, S. Vishwanathan, and R. Garnett, Eds., vol. 30, Curran Associates, Inc., 2017.

[2] J. Devlin, M.-W. Chang, K. Lee, and K. Toutanova, "BERT: Pre-training of deep bidirectional transformers for language understanding," in *Proceedings of the 2019 Conference of the North American Chapter of the Association for Computational Linguistics: Human Language Technologies*, Volume 1 (Long and Short Papers), Association for Computational Linguistics, pp. 4171–4186, 2019.

[3] R. Thoppilan, D. D. Freitas, J. Hall, N. Shazeer, A. Kulshreshtha, H.-T. Cheng, A. Jin, et al., "LaMDA: Language models for dialog applications," 2022.

[4] T. Brown, B. Mann, N. Ryder, M. Subbiah, J. D. Kaplan, P. Dhariwal, A. Neelakantan, et al., "Language models are few-shot learners," in *Advances in Neural Information Processing Systems*, H. Larochelle, M. Ranzato, R. Hadsell, M. Balcan, and H. Lin, Eds., vol. 33, Curran Associates, Inc., pp. 1877-1901, 2020.

[5] J. Wei, Y. Tay, R. Bommasani, C. Raffel, B. Zoph, S. Borgeaud, D. Yogatama, et al., "Emergent abilities of large language models," 2022.

[6] P. Törnberg, "ChatGPT-4 outperforms experts and crowd workers in annotating political twitter messages with zero-shot learning," 2023.

[7] Y. Li, D. Choi, J. Chung, N. Kushman, J. Schrittwieser, R. Leblond, T. Eccles, et al., "Competition-level code generation with alphacode," *Science*, 378(6624):1092–1097, 2022.

[8] V. G. Cerf, "Large language models," *Communications of the ACM*, vol. 66, no. 8, p. 7, 2023.

[9] Z. Ji, N. Lee, R. Frieske, T. Yu, D. Su, Y. Xu, E. Ishii, Y. J. Bang, A. Madotto, and P. Fung, "Survey of Hallucination in Natural Language Generation," *ACM Computing Surveys*, Vol. 55, Issue 12, Article 248, 38 pages, 2023.

[10] Y. Liang, C. Wu, T. Song, W. Wu, Y. Xia, Y. Liu, Y. Ou, S. Lu, L. Ji, S. Mao, Y. Wang, L. Shou, M. Gong, and N. Duan, "Taskmatrix.ai: Completing tasks by connecting foundation models with millions of APIs," 2023.

[11] J. Maynez, S. Narayan, B. Bohnet, and R. McDonald, "On faithfulness and factuality in abstractive summarization," in *Proceedings of the 58th Annual Meeting of the Association for Computational Linguistics*, pp. 1906–1919, 2020.

[12] P. Budzianowski, T.-H. Wen, B.-H. Tseng, I. Casanueva, S. Ultes, O. Ramadan, and M. Gašić, "Multiwoz – a large-scale multi-domain wizard-of-oz dataset for task-oriented dialogue modelling," 2020.

[13] Q. Jia, Y. Liu, S. Ren, and K. Q. Zhu, "Taxonomy of abstractive dialogue summarization: Scenarios, approaches, and future directions," *ACM Computing Surveys*, vol. 56, no. 3, 2023.

[14] X. Zang, A. Rastogi, S. Sunkara, R. Gupta, J. Zhang, and J. Chen, "MultiWOZ 2.2: A Dialogue Dataset with Additional Annotation Corrections and State Tracking Baselines", in *Proceedings of the 2nd Workshop on Natural Language Processing for Conversational AI, Association for Computational Linguistics*, pp. 109–117, 2020.

[15] L. Zhao, F. Zheng, K. He, W. Zeng, Y. Lei, H. Jiang, W. Wu, W. Xu, J. Guo, and F. Meng, "TODSum: Task-oriented dialogue summarization with state tracking," arXiv preprint arXiv:2110.12680, 2021.

[16] S. R. Bowman, "Eight things to know about large language models," 2023

[17] K. S. Yim, *From Experiment To Design – Fault Characterization and Detection in Parallel Computer Systems Using Computational Accelerators*, Ph.D. Dissertation, University of Illinois at Urbana-Champaign, May 2013.

[18] S. Mehri, J. Choi, L. F. D'Haro, J. Deriu, M. Eskenazi, M. Gasic, et al., "Report from the NSF Future Directions Workshop on Automatic Evaluation of Dialog: Research Directions and Challenges," arXiv:2203.10012, 2022.

[19] B. Byrne, K. Krishnamoorthi, C. Sankar, A. Neelakantan, D. Duckworth, S. Yavuz, B. Goodrich, A. Dubey, A. Cedilnik, and K.-Y. Kim, "Taskmaster-1: Toward a Realistic and Diverse Dialog Dataset," arXiv 1909.05358, 2019.

[20] Y. Zhang, Y. Li, L. Cui, D. Cai, L. Liu, T. Fu, X. Huang, E. Zhao, et al., "Siren's Song in the AI Ocean: A Survey on Hallucination in Large Language Models," arXiv:2309.01219, 2023.

[21] L. Huang, W. Yu, W. Ma, W. Zhong, Z. Feng, H. Wang, Q. Chen, W. Peng, et al., "A Survey on Hallucination in Large Language Models: Principles, Taxonomy, Challenges, and Open Questions," arXiv:2311.05232, 2023.

[22] K. S. Yim, I. Firman, A. Coimbra, R. Berry, M. Ionut-Andreica, M. Reutov, G. Taubman, C. Kuang, M. Oh, S. Ganov, and K. Desineni, "Human-In-The-Loop Voice Automation System," US Patent Application, No. 18,013,083, August 2023.

[23] W. Fish, *Perception, Hallucination, and Illusion*, Oxford University Press, 2009.

[24] J. D. Blom, *A Dictionary of Hallucinations*, Springer, 2010.

[25] K. Filippova, "Controlled hallucinations: Learning to generate faithfully from noisy data," in *Proceedings of the 2020 Conference on Empirical Methods in Natural Language Processing: Findings*, pp. 864-870, 2020.

[26] X. Liu, A. Eshghi, P. Swietojanski, and V. Rieser, "Benchmarking Natural Language Understanding Services for building Conversational Agents," arXiv:1903.05566, 2019.

[27] J. Maynez, S. Narayan, B. Bohnet, and R. McDonald, "On faithfulness and factuality in abstractive summarization," in *Proceedings of the 58th Annual Meeting of the Association for Computational Linguistics*, 2020.

[28] C. Zhou, G. Neubig, J. Gu, M. Diab, F. Guzmán, L. Zettlemoyer, and M. Ghazvininejad, "Detecting hallucinated content in conditional neural sequence generation," in *Findings of the Association for Computational Linguistics (ACL-IJCNLP)*, Association for Computational Linguistics, pp. 1393–1404, 2021.

[29] A. Köksal, R. Aksitov, and C.-C. Chang, "Hallucination Augmented Recitations for Language Models," arXiv: 2311.07424, 2023.

[30] A. Pagnoni, V. Balachandran, and Y. Tsvetkov, "Understanding factuality in abstractive summarization with FRANK: A benchmark for factuality metrics," in *Proceedings of the 2021 Conference of the North American Chapter of the Association for Computational Linguistics: Human Language Technologies (NAACL-HLT)*, pp. 4812–4829, 2021.

[31] Y. Li, K. Yao, L. Qin, W. Che, X. Li, and T. Liu, "Slot-consistent NLG for task-oriented dialogue systems with iterative rectification network," in *Proceedings of the Annual Meeting of the Association for Computational Linguistics*, pp. 97–106, 2020.

[32] S. Santhanam, B. Hedayatnia, S. Gella, A. Padmakumar, S. Kim, Y. Liu, and D. Hakkani-Tur, "Rome was built in 1776: A case study on factual correctness in knowledge-grounded response generation," arXiv preprint arXiv:2110.05456(2021), 2021.

[33] K. Papineni, S. Roukos, T. Ward, and W.-J. Zhu, "BLEU: a method for automatic evaluation of machine translation," in *Proceedings of the Annual Meeting on Association for Computational Linguistics (ACL)*, pp. 311–318, 2002.

[34] C.-Y. Lin, "ROUGE: A Package for Automatic Evaluation of Summaries", in *Proceedings of the Annual Meeting on Association for Computational Linguistics (ACL)*, pp. 74–81, 2004.

[35] K. Lee, O. Firat, A. Agarwal, C. Fannjiang, and D. Sussillo, "Hallucinations in Neural Machine Translation," in *Interpretability and Robustness in Audio, Speech, and Language (IRASL) Workshop, Conference on Neural Information Processing Systems (NeurIPS)*, 2018.

[36] K. Shuster, S. Poff, M. Chen, D. Kiela, and J. Weston, "Retrieval augmentation reduces hallucination in conversation," in *Findings of the Association for Computational Linguistics (EMNLP)*, Association for Computational Linguistics, pp. 3784–3803, 2021.

[37] T. Yu, Z. Liu, and P. Fung, "AdaptSum: Towards low-resource domain adaptation for abstractive summarization," in

*Proceedings of the Conference of the North American Chapter of the Association for Computational Linguistics: Human Language Technologies (NAACL-HLT)*, pp. 5892–5904, 2021.

[38] Z. Wang, X. Wang, B. An, D. Yu, and C. Chen, "Towards faithful neural table-to-text generation with content-matching constraints," in *Proceedings of the Annual Meeting of the Association for Computational Linguistics*, 2020.

[39] B. Goodrich, V. Rao, P. J. Liu, and M. Saleh, "Assessing the factual accuracy of generated text," in *Proceedings of the 25th ACM SIGKDD International Conference on Knowledge Discovery and Data Mining*, pp. 166–175, 2019.

[40] E. Durmus, H. He, and M. Diab, "FEQA: A question answering evaluation framework for faithfulness assessment in abstractive summarization," in *Proceedings of the Annual Meeting of the ACL*, pp. 5055–5070, 2020.

[41] O. Honovich, L. Choshen, R. Aharoni, E. Neeman, I. Szpektor, and O. Abend, "Q2: Evaluating factual consistency in knowledge-grounded dialogues via question generation and question answering," in *Proceedings of the Conference on Empirical Methods in Natural Language Processing*, pp. 7856–7870, 2021.

[42] C. Rebuffel, T. Scialom, L. Soulier, B. Piwowarski, S. Lamprier, J. Staiano, G. Scoutheeten, and P. Gallinari, "Data-QuestEval: A reference-less metric for data-to-text semantic evaluation," in *Proceedings of the Conference on Empirical Methods in Natural Language Processing*, 2021.

[43] T. Falke, L. F. R. Ribeiro, P. A. Utama, I. Dagan, and I. Gurevych, "Ranking generated summaries by correctness: An interesting but challenging application for natural language inference," in *Proceedings of the Annual Meeting of the Association for Computational Linguistics*, pp. 2214–2220, 2019.

[44] C. Zhou, G. Neubig, J. Gu, M. Diab, F. Guzmán, L. Zettlemoyer, and M. Ghazvininejad, "Detecting hallucinated content in conditional neural sequence generation," in *Findings of the Association for Computational Linguistics (CL-IJCNLP)*, Association for Computational Linguistics, pp. 1393–1404, 2021.

[45] Y. Wang, S. Mishra, P. Alipoormolabashi, Y. Kordi, A. Mirzaei, A. Naik, A. Ashok, *et al.*, "Super-NaturalInstructions: Generalization via Declarative Instructions on 1600+ NLP Tasks," in *Proceedings of the Conference on Empirical Methods in Natural Language Processing*, pp 5085–5109, 2022.

[46] H. W. Chung, L. Hou, S. Longpre, B. Zoph, Y. Tay, W. Fedus, Y. Li, X. Wang, M. Dehghani, *et al.*, "Scaling Instruction-Finetuned Language Models," *arXiv:2210.11416*, 2022.

[47] V. Sanh, A. Webson, C. Raffel, S. Bach, L. Sutawika, Z. Alyafeai, A. Chaffin, A. Stiegler, A. Raja, M. Dey, *et al.*, "Multitask Prompted Training Enables Zero-Shot Task Generalization," in *Proceedings of the International Conference on Learning Representations*, 2022.